\documentclass{ws-jai}
\usepackage[flushleft]{threeparttable}
\usepackage{hyperref}
\usepackage{tikz}
\usepackage{caption}
\usepackage{subcaption}
\usepackage{color,soul}
\usepackage{xcolor}
\usetikzlibrary{shapes, arrows}
\tikzstyle{terminator} = [rectangle, draw, text centered, rounded corners, minimum height=2em]
\tikzstyle{process} = [rectangle, draw, text centered, minimum height=2em, align=center]
\tikzstyle{decision} = [diamond, draw, text centered, minimum height=2em]
\tikzstyle{data}=[trapezium, draw, text centered, trapezium left angle=60, trapezium right angle=120, minimum height=2em]
\tikzstyle{connector} = [draw, -latex']

\definecolor{myblue1}{RGB}{120, 210, 255}
\definecolor{myyellow1}{RGB}{255, 255, 185}

\begin{document}

\catchline{}{}{}{}{} % Publisher's Area please ignore

\markboth{V. MacKay et al.}{Low-cost, Low-loss, Ultra-wideband Compact Feed for Interferometric Radio Telescopes}

\title{Low-cost, Low-loss, Ultra-wideband Compact Feed for  Interferometric Radio Telescopes}

\author{Vincent MacKay$^{1,2,6}$, Mark Lai$^4$, Peter Shmerko$^4$, Dallas Wulf$^5$, Leonid Belostotski$^4$, Keith Vanderlinde$^{2,3,1}$}

\address{
$^{1}$Department of Physics, University of Toronto, 60 St George St, Toronto, ON, M5S 1A7, Canada\\
$^{2}$Dunlap Institute for Astronomy and Astrophysics, University of Toronto, 50 St George St, Toronto, ON, M5S 3H4, Canada\\
$^{3}$David A. Dunlap Department of Astronomy \& Astrophysics, University of Toronto, 50 St George St, Toronto, ON, M5S 3H4, Canada\\
$^{4}$Department of Electrical and Software Engineering, University of Calgary, 2500 University Dr., Calgary, AB, T2N 1N4\\
$^{5}$Department of Physics, McGill University, 3600 rue University, Montréal, QC, H3A 2T8, Canada\\
$^{6}$vincent.mackay@mail.utoronto.ca
}

\maketitle

\corres{$^{1}$Corresponding author.}

\begin{history}
\received{(to be inserted by publisher)};
\revised{(to be inserted by publisher)};
\accepted{(to be inserted by publisher)};
\end{history}

\begin{abstract}
We have developed, built, and tested a new feed design for interferometric radio telescopes with ``large-$N$, small-$D$'' designs. Those arrays require low-cost and low-complexity feeds for mass production on reasonable timescales and budgets, and also require those feeds to be compact to minimize obstruction of the dishes, along with having ultra wide bands of operation for most current and future science goals. The feed presented in this paper modifies the exponentially tapered slot antenna (Vivaldi) and quad-ridged flared horn antenna designs by having an oversized backshort, a novel method of maintaining a small size that is well-suited for deeper dishes ($f/D\leq 0.25$). It is made of laser cut aluminum and printed circuit boards, such that it is inexpensive ($\lesssim$\,75\,USD per feed in large-scale production) and quick to build; it has a 5:1 frequency ratio, and its size is approximately a third of its longest operating wavelength. We present the science and engineering constraints that went into design decisions, the development and optimization process, and the simulated performance. A version of this feed design was optimized and built for the Canadian Hydrogen Observatory and Radio-transient Detector (CHORD) prototypes. When simulated on CHORD's very deep dishes ($f/D=0.21$) and with CHORD's custom first stage amplifiers, the on-sky system temperature $T_\mathrm{sys}$ of the complete receiving system from dish to digitizer remains below 30\,K over most of the 0.3--1.5\,GHz band, and maintains an aperture efficiency $\eta_\mathrm{A}$ between 0.4 and 0.6. The entire receiving chain operates at ambient temperature. The feed is designed to slightly under-illuminate the CHORD dishes, in order to minimize coupling between array elements and spillover. 
\end{abstract}


\section{Introduction}
\label{sec:intro}

Modern advances in telecommunication technology and improvements in computing capabilities have enabled radio astronomers to build ``large-$N$, small-$D$'' observatories---where $N$ is the number of dishes, and $D$, their diameter---such as the DSA-2000 HIRAX, HERA, or PUMA, which is set to have $N=$ 32,000 \citep{HERA_overview,DSA_overview,puma,HIRAX_overview}. This requires the design of low-cost and low-complexity signal chains, from dish to digitizer, so that arrays with $N\gtrsim\mathcal{O}(1000)$ can be realized for a low cost and within ambitious timelines. The feeds also need to be compact, since the small diameter of the dishes means that they are easily obstructed. An important science goal for those observatories is 21\,cm intensity mapping, which additionally requires very low-noise, dual-polarized, ultra-wideband receivers---see \citealt{Furlanetto_review}, \citealt{Pritchard_review}, and \citealt{Liu_review} for in-depth reviews of the science and constraints.

We present a feed design that meets those challenges. It follows the design of an exponentially tapered slot antenna (Vivaldi), and is also inspired by recent literature on the quad-ridge flared horn (QRFH) feed designs \citep{akgiray_circular,akgiray_thesis,flygare_thesis}. One of its novel feature is having the cut behind the exponential taper---the \textit{backshort}, see \autoref{fig:petal_diagram}---be oversized, which extends its band towards the lower frequencies while maintaining the overall size of the feed near or below $0.4\times0.3\lambda_0$ (width $\times$ length) in size for each polarization, where $\lambda_0 = 1$\,m is the longest wavelength at which it operates, at a 5:1 frequency ratio. Many degrees of freedom in the feed and balun geometries allow for a fine-tuning of the impedance and beam shape for the specific requirements of a given array, and we summarize an optimization routine that can be used to adapt the feed to such requirements. The feed is laser cut out of 3.175\,mm-thick aluminum, and the baluns are printed on 0.8\,mm-thick low-loss ($D_k\sim3.5$) substrates. The manufacture process allows the feed to be quickly mass-produced below the 75\,USD price point, and the losses in materials result in only 1--4\,K increase in system temperature over the band (0.014--0.058\,dB loss). 

We show an implementation of this feed design for the Canadian Hydrogen Observatory and Radio-transient Detector (CHORD), an array of 512 closely packed 6\,m prime-focus dishes set to be completed by the mid-2020's in British Columbia, Canada, along with two ``outrigger'' stations for VLBI, on the US East and West coasts. CHORD will aim at mapping neutral hydrogen at redshifts $z<3.7$, detecting and precisely locating fast radio bursts (FRBs) and other radio transients, probing cosmic magnetism, and more \citep{CHORD_overview}. Over most of its 0.3--1.5\,GHz frequency coverage, the CHORD feed exhibits on-sky system temperature $T_\mathrm{sys} \lesssim$ 30\,K at ambient temperature (taken as 300\,K throughout), and maintains an aperture efficiency $\eta_\mathrm{A}$ between 0.4 and 0.6 when mounted on CHORD's very deep $f/D = 0.21$ dishes. The feed's dimensions are $0.4\times 0.3\lambda_0$ (width $\times$ length), such that dish blockage is minimized. The manufacture of all its parts---laser cut aluminum, and PCBs---is computer-controlled, and the materials are inexpensive, so that it can be mass produced at low cost. Approximately 20 minutes of manual labor is required for the assembly of one feed.

In \autoref{sec:design}, we present the considerations and constraints that have led to the design decisions. In \autoref{sec:development}, we summarize the modeling, optimization, and manufacture process. We present simulated and measured performances in \autoref{sec:performance}. Lastly, in \autoref{sec:CHORD}, we describe CHORD and its specific constraints, and we present the performance of the feed when used with the CHORD dishes and custom LNAs.

\section{Design choice, considerations, and constraints}
\label{sec:design}

\subsection{A compact feed}
\label{subsec:miniaturizing}

The design chosen for the feed is that of an exponentially tapered slot antenna, commonly known as a ``Vivaldi'' feed, inspired by the one used for HERA \citep{vivaldi_original,hera_feeds}. The Vivaldi design is ultra-wideband, easy to manufacture, can be made to be dual-polarized, and has many degrees of freedom to optimize for a desired feed size, impedance and beam shape.

%One of the most common approaches to minimize the size of a Vivaldi feed involves printing each side of the exponential taper on opposite faces of a substrate---the so-called \textit{antipodal} design \citep{Antipodal_1,Antipodal_2,Antipodal_3}. However, substrates used to support PBC antennas are either very lossy, or very expensive, so we use self-supporting ground planes, which is incompatible with an antipodal design. Another common method of miniaturization is to add resistive loads to improve the impedance match over a wider band, but we also had to discard this approach due to the increased losses \citep{vivaldi_lumped_resistor_1,vivaldi_lumped_resistor_2,vivaldi_lumped_resistor_3,vivaldi_lumped_resistor_4}.

The degree of freedom that we chose to precisely tune in order to keep the feed compact is the shape of its outline, as this can be done without increasing cost, complexity, or losses. For example, it has been shown that narrowing each polarization plane by cutting a second exponential taper along their outer edges, and, in the space thus created, adding carefully shaped stubs where longer wavelength currents can circulate, allows for the feed to radiate lower frequencies without increasing its dimensions \citep{vivaldi_stubs_1,vivaldi_stubs_2,vivaldi_stubs_3}. This approach was explored for our uses, but we found that some out-of-phase radiation happened through the added stubs at higher frequencies, suppressing the forward gain and increasing the size of the sidelobes, especially when the feed was used inside deeper dishes $f/D\leq 0.25$---see \autoref{subsec:CHORD_crosstalk} for discussion about dish depth). We instead maintained a small size by engineering the feed outline through a novel method: by significantly increasing the size of the backshort to approximately $0.15\lambda_0$. The overall size of the feed presented in \autoref{sec:CHORD} is kept at $0.4\times 0.3\lambda_0$,  below the $\sim 0.5 \lambda_0$ length of a standard Vivaldi feed. This design choice significantly widens the beam, making the feed well-suited for deep dishes, but inadequate for shallower ones---that trade-off is presented in \autoref{sec:performance}, and illustrated in \autoref{fig:eta_A_T_sys}.

\subsection{Comparison with other feeds}
\label{subsec:QRFH}
The self-supporting Vivaldi design is equivalent to an open boundary QRFH design, which is a broadband feed design widely used in radio astronomy on observatories such as the SKA \citep{akgiray_circular,SKA_Band_1,SKA_overview}. Exhaustive reviews of the QRFH family of feeds in the context of radio astronomy are presented in \citealt{akgiray_thesis} and \citealt{flygare_thesis}. The latter notes that deeper dishes, with $f/D\lesssim 0.3$, are harder to illuminate over wide bands, especially using low-cost, low-complexity feeds that cannot make use of intricate flared horns, choke rings, or dielectric loads to shape their beam. In \autoref{sec:performance}, we show that our design does achieve good illumination ($\eta_\mathrm{A}\sim$ 0.5--0.6) on deep dishes ($f/D\leq 0.25$), while indeed being inexpensive and simple to manufacture. In \autoref{subsec:CHORD_crosstalk}, we discuss the benefits of an under-illuminated dish for CHORD's science goals.

Note that additional degrees of freedom in the feed outline have been explored in the QRFH literature. For instance, any active segment of the outline can be replaced by a spline, as shown in \citealt{spline_1} and \citealt{spline_2}, and alternate backshort geometries are presented in \citealt{backshort}. Both of these methods open a broader parameter space to potentially reach more a desirable impedance and beam shape. These options have not been explored yet for the feed design proposed in this paper.

In \autoref{tab:feed_comparison}, we compare the performance and cost of the CHORD feed with that of other ``large-$N$, small-$D$'' observatories. The DSA-2000 feed is a QRFH \citep{DSA_overview}; the CHIME feed is a modified four-square antenna \citep{chime_feed, meiling_in_prep}, an updated version with active baluns and different substrates is set to be used on HIRAX \citep{HIRAX_overview}; the HERA feed is a Vivaldi \citep{HERA_overview,hera_feeds}. The CHIME feed's $\eta_\mathrm{A}$ is not presented as it is not a useful metric for cylindrical reflectors. Costs were obtained through private correspondence with members from each collaboration; they include both polarizations, but exclude active parts as well as manufacturing and testing. Additionally, they are only estimated to be correct within $\pm 30\%$ since the cost of materials and manufacturing varies appreciably as a function of country, year, and specific circumstances such as access to machine shops or material surplus. The higher cost of the HERA feeds can be explained by their significantly larger sizes, as HERA probes longer wavelengths ($\lambda_0=6$\,m).

\begin{table}[h!]\centering
\begin{threeparttable}[b]
        \caption{Feeds used on various current and proposed observatories. $T_\mathrm{sys}$ and $\eta_\mathrm{A}$ were band-averaged. The cost is per feed, and estimated for large-scale production.}
\begin{tabular}{lllllll}\hline
Observatory       & Band (GHz) & $\lambda_0$ (m) & $T_\mathrm{sys}$ (K) & $\eta_\mathrm{A}$ & W $\times$ L ($\lambda_0$) & Cost (USD)  \\\hline
DSA-2000                   & 0.7--2 (2.86:1) &  0.43 & 25 & 0.7      & $1.16 \times 0.93$ & \$300$^\dagger$  \\
CHIME                  &  0.4--0.8 (2:1) &  0.75  & 40 & ---       & $0.36 \times 0.2$ & \$80$^\ddagger$ \\
HIRAX                    & 0.4--0.8 (2:1) &  0.75  & 50 & 0.5      & $0.36 \times 0.2$ & \$60$^\dagger$  \\
HERA                       & 0.05--0.25 (5:1) &  6   & 50 & 0.5      & $0.3 \times 0.26$ & \$1400$^{\ddagger}$ \\
CHORD (this work)          & 0.3--1.5 (5:1) &  1  & 30 & 0.55     & $0.4 \times 0.3 $ & \$75$^\dagger$  \\
\hline
\end{tabular}\label{tab:feed_comparison}
\begin{tablenotes}[para]
    \footnotesize
     $^\dagger$Projected cost. $^{\ddagger}$Realized cost.
\end{tablenotes}
\end{threeparttable}
\end{table}

\section{Development}
\label{sec:development}

\subsection{Modeling the feed}
\label{subsec:modeling}
\begin{figure}[h!]
    \centering
    \includegraphics[width=1.0\textwidth]{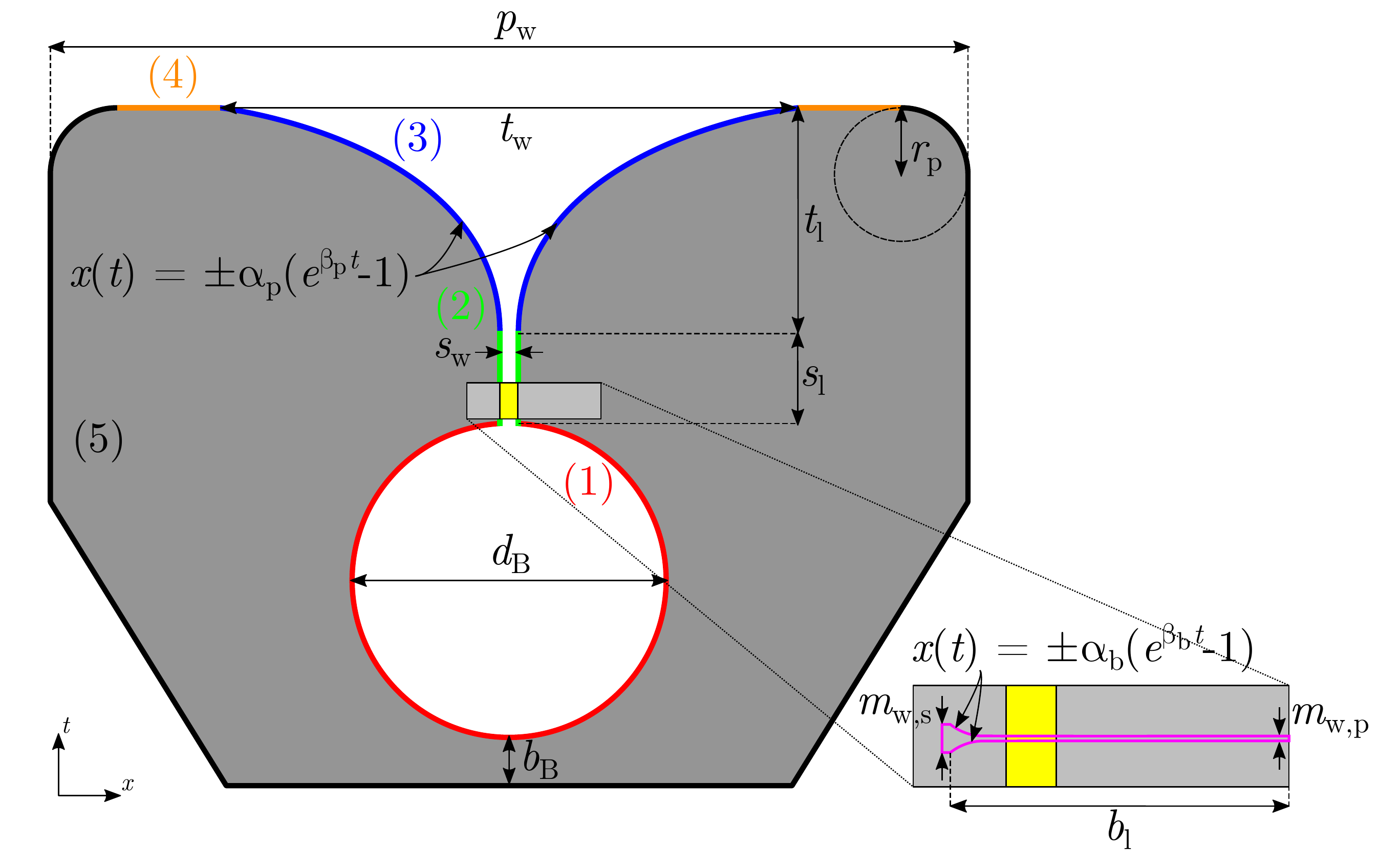}
    \caption{Diagram of a feed petal (left) and balun (bottom right). The labeled parts are (1) the backshort, (2) the untapered slot, (3) the exponential taper and (4) its straight extension, and (5) the shape rest of the outline. The pink curve on the balun diagram shows the outline of the exponentially tapered microstrip printed on the other side. The values for the parameters are presented in \autoref{tab:dimensions}.}
    \label{fig:petal_diagram}
\end{figure}

\begin{figure}[h!]
     \centering
     \begin{subfigure}[b]{0.61\textwidth}
         \centering
         \includegraphics[width=\textwidth]{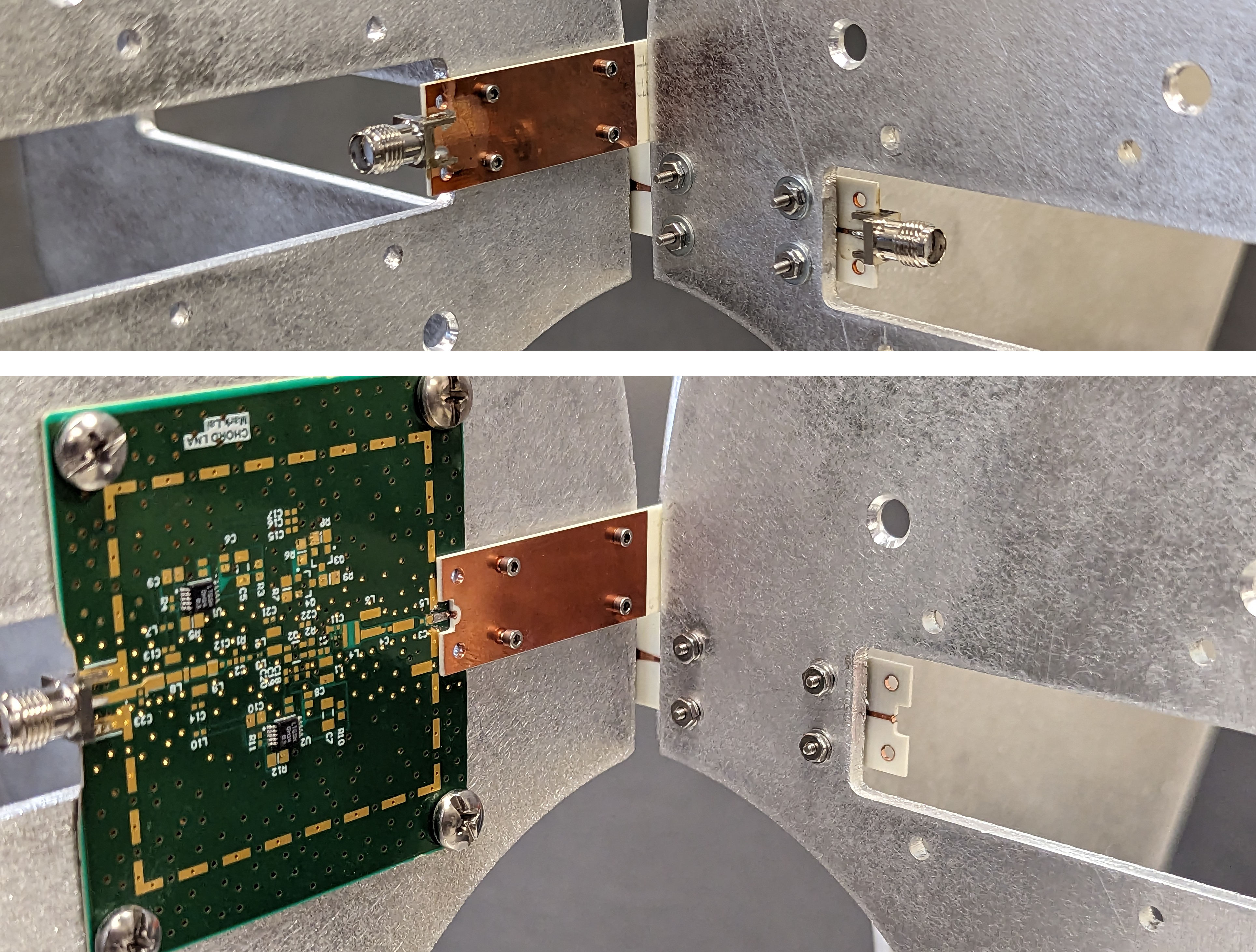}
         \caption{}
         \label{subfig:baluns_on_feed}
     \end{subfigure}
     \hfill
     \begin{subfigure}[b]{0.34\textwidth}
         \centering
         \includegraphics[width=\textwidth]{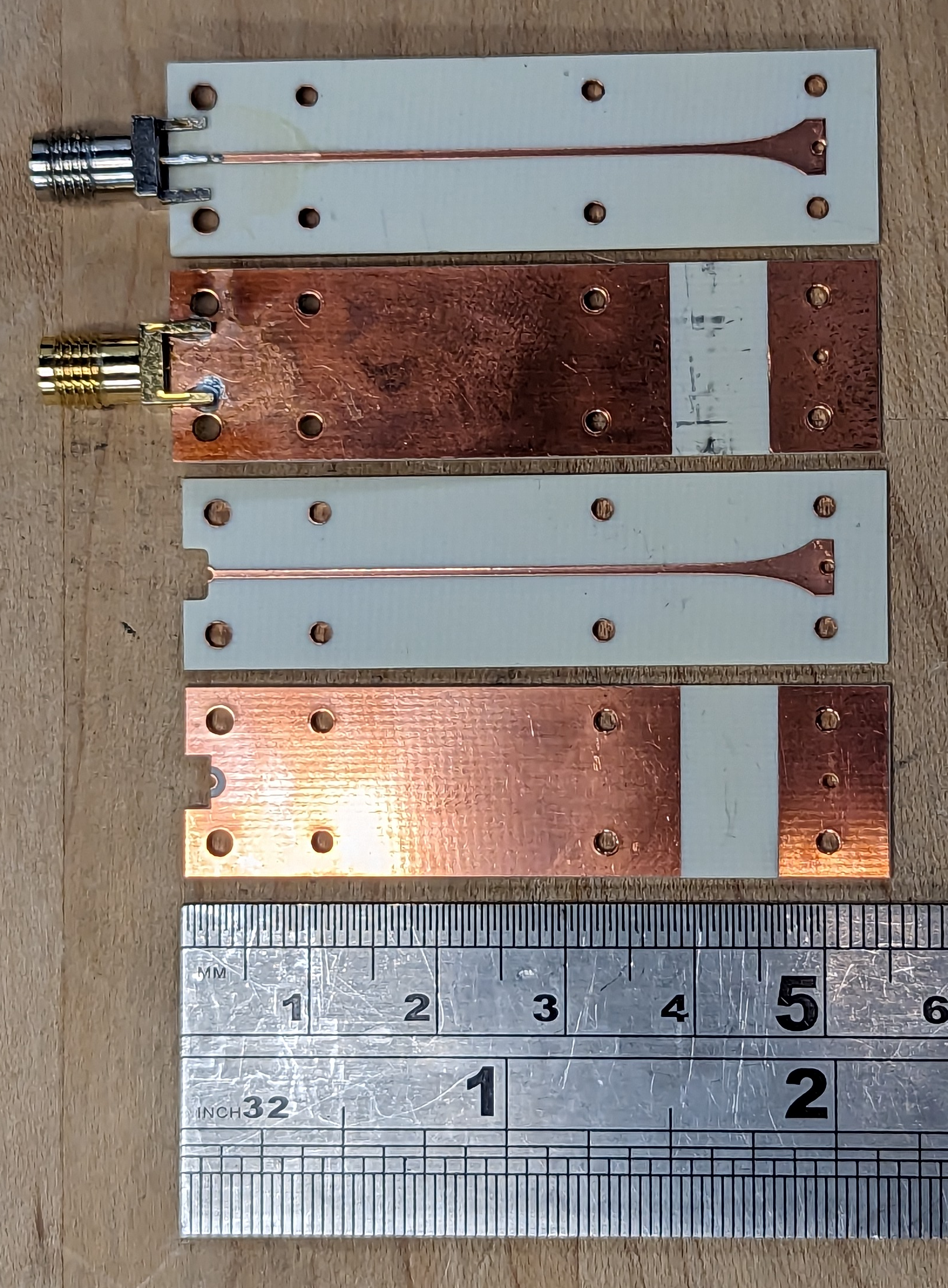}
         \caption{}
         \label{subfig:baluns_naked}
     \end{subfigure}
        \caption{(a) (Top) Two microstrip baluns terminated by SMA connectors, as installed on each polarization petals of the feed. The balun closest to the backshort is labeled ``polarization 0,'' and the one closest to the exponential taper is labeled ``polarization 1.'' Not shown: the short to the petal, on the other end, which is done through an off-the-shelf connection pin. (Bottom) Alternate microstrip baluns without SMA connectors, with a custom CHORD amplifier on one plane. A box covering the amplification circuit was removed for the picture. (b) The two different types of microstrip baluns, before being installed on the feed.}
        \label{fig:baluns_fig}
\end{figure}

The feed was modeled and simulated using CST Microwave Studio. It is composed of two perpendicular polarization planes, that we call the \textit{petals}, each equipped with a small microstrip balun, and one circular plane at the back---the \textit{backplane}---that mostly serves mounting purposes. Apart from the locations of the baluns, which must be offset along the untapered slot (see \autoref{subfig:baluns_on_feed}), the petals are identical. The offset in the positions of the baluns results in a negligible difference in the impedance matching between the two polarizations (see \autoref{fig:reflection}). That difference could be corrected for by optimizing the dimensions of the baluns separately, but we found that it was not necessary to reach standard performance requirements. The petals are extruded profiles of a closed planar curve, which is presented in \autoref{fig:petal_diagram} and defined by the following features:
\begin{enumerate}
    \item a large circular cut at the back, called the \textit{backshort}, providing a control over impedance,
    \item a short untapered slot, where the baluns are attached,
    \item an exponential taper, defined by $x(t) = \alpha_\mathrm{p} \left(e^{\beta_\mathrm{p} t}-1\right)$,
    \item a straight extension of the taper,
    \item the rest of the contour, which is defined entirely for structural reasons, and to leave space for first-stage amplification.
\end{enumerate}
Note that in Vivaldi and QRFH feeds, the backshort typically has the added function of preventing radiation towards the back. As our design's backshort is oversized, this effect is attenuated, and the feed radiates an appreciable amount of power in the back hemisphere. This result is discussed in \autoref{sec:performance}.

\begin{figure}[h!]
     \centering
     \begin{subfigure}[b]{0.3\textwidth}
         \centering
         \includegraphics[width=\textwidth]{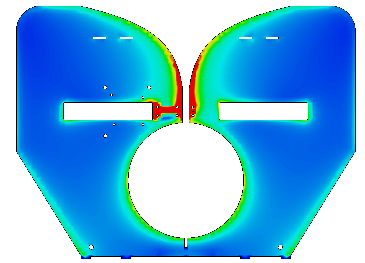}
         \caption{}
         \label{subfig:currents0.3}
     \end{subfigure}
     \hfill
     \begin{subfigure}[b]{0.3\textwidth}
         \centering
         \includegraphics[width=\textwidth]{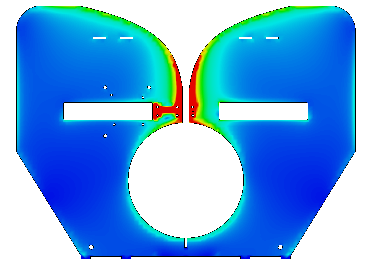}
         \caption{}
         \label{subfig:currents0.9}
     \end{subfigure}
     \hfill
     \begin{subfigure}[b]{0.3\textwidth}
         \centering
         \includegraphics[width=\textwidth]{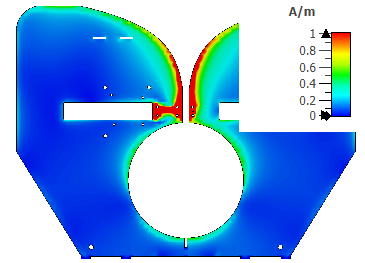}
         \caption{}
         \label{subfig:currents1.5}
     \end{subfigure}\\
     \begin{subfigure}[b]{0.3\textwidth}
         \centering
         \includegraphics[width=\textwidth]{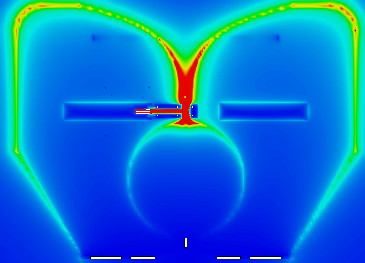}
         \caption{}
         \label{subfig:efield_below_balun0.3}
     \end{subfigure}
     \hfill
     \begin{subfigure}[b]{0.3\textwidth}
         \centering
         \includegraphics[width=\textwidth]{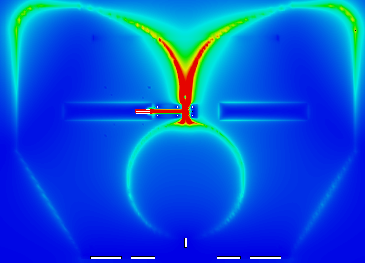}
         \caption{}
         \label{subfig:efield_below_balun0.9}
     \end{subfigure}
     \hfill
     \begin{subfigure}[b]{0.3\textwidth}
         \centering
         \includegraphics[width=\textwidth]{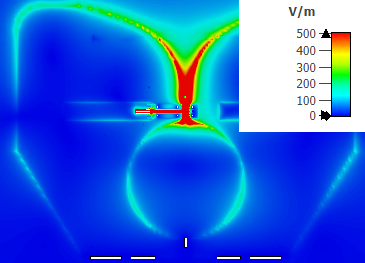}
         \caption{}
         \label{subfig:efield_below_balun1.5}
     \end{subfigure}\\
     \begin{subfigure}[b]{0.3\textwidth}
         \centering
         \includegraphics[width=\textwidth]{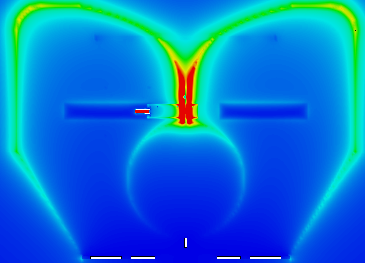}
         \caption{}
         \label{subfig:efield_above_balun0.3}
     \end{subfigure}
     \hfill
     \begin{subfigure}[b]{0.3\textwidth}
         \centering
         \includegraphics[width=\textwidth]{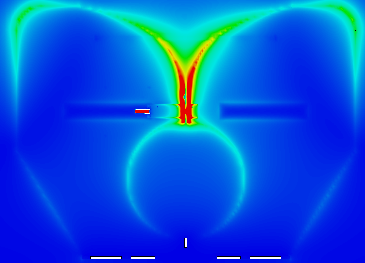}
         \caption{}
         \label{subfig:efield_above_balun0.9}
     \end{subfigure}
     \hfill
     \begin{subfigure}[b]{0.3\textwidth}
         \centering
         \includegraphics[width=\textwidth]{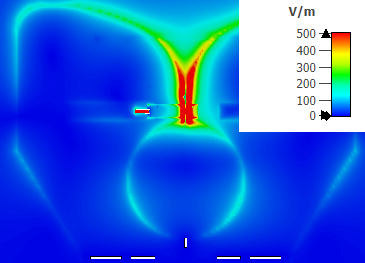}
         \caption{}
         \label{subfig:efield_above_balun1.5}
     \end{subfigure}\\
        \caption{First row: RMS of the current density on the excited petal, as simulated using CST, at (a) 0.3\,GHz, (b) 0.9\,GHz, and (c) 1.5\,GHz. Second row: at the same frequencies, RMS of the E-field on a plane that lies 0.4\,mm above the petal and parallel to it, below the balun. Note the high field density between the balun and the petal. Third row: at the same frequencies, RMS of the E-field on a plane that lies 0.1\,mm above the balun. Note the absence of field along the balun's microstrip, indicating that the grounded outer layer of the balun works as an effective shield.}
        \label{fig:currents}
\end{figure}

\begin{figure}[h!]
    \centering
    \includegraphics[width=0.7\textwidth]{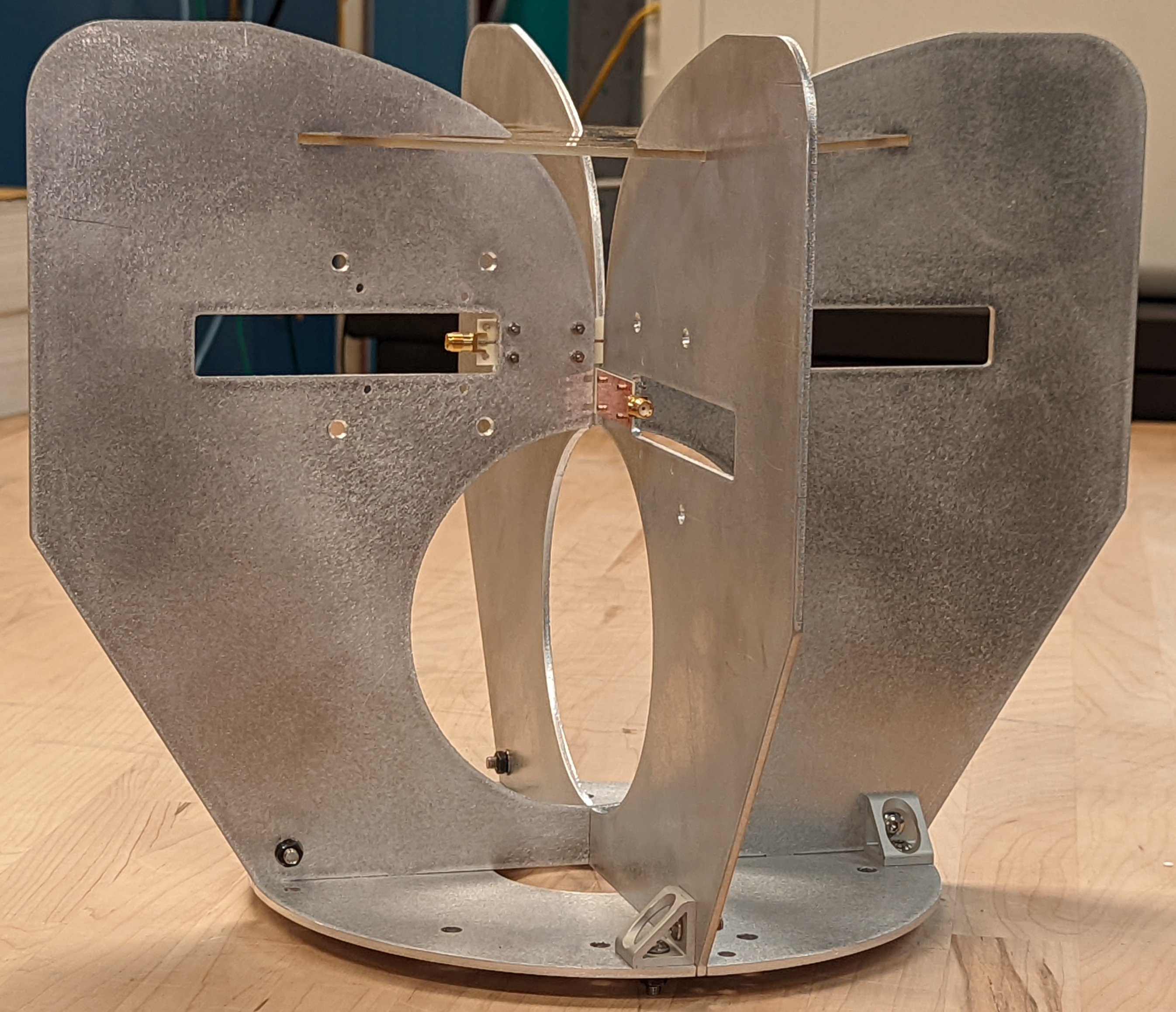}
    \caption{Completed feed, including the microstrip baluns terminated by SMA connectors. There are cuts in the petals that are added to accommodate screws, cables, connectors, and first stage amplification. There is also a stabilizer piece, inserted in the exponential taper, made of polycarbonate. This version of the feed design is the one optimized for CHORD, as presented in \autoref{sec:CHORD}.}
    \label{fig:feed_alone}
\end{figure}

Additional cuts are made inside that profile to accommodate screws, connectors, cables, and first-stage amplification, as can be seen in \autoref{fig:feed_alone}. Those cuts do not affect the performance of the feed, as most of the electrical activity happens near the inside edges of the petals, as can be seen in \autoref{fig:currents}. The relevant free parameters of that design are the diameter of the backshort, the length and width of the untapered slot, the length and maximum opening of the exponential taper, the taper rate parameter $\beta_\mathrm{p}$, and the total width of the petals, which include the taper's straight extension and the blended corners. The taper scale parameter $\alpha_\mathrm{p}$ is not free, as it is uniquely determined by the taper's length and width. 

The petals and backplane's thickness is fixed at 3.175\,mm (1/8\,in.), which is a standard metal thickness in North America, and is thick enough that the feed is self-supporting without needing to be printed on a substrate. The material chosen is aluminum as it has a high conductivity, it is light, inexpensive, and its oxidization has negligible effects on performances at low frequencies.

\begin{table}\centering
    \begin{threeparttable}[b]
        \caption{Design parameters for the feed, illustrated in \autoref{fig:petal_diagram}, optimized for the CHORD reflector and science goals.}
        \begin{tabular}{lll}
        \hline
        Parameter                                             & Symbol        & Value \\ \hline
        Petal backshort diameter                              & $b_\mathrm{d}$         & 65 mm      \\
        Petal untapered slot width                            & $s_\mathrm{w}$         & 7.5 mm       \\
        Petal untapered slot length                           & $s_\mathrm{l}$         & 37 mm      \\
        Petal taper length                                    & $t_\mathrm{l}$         & 93 mm      \\
        Petal taper opening width                             & $t_\mathrm{w}$         & 238 mm        \\
        Petal taper scale parameter (fixed)                   & $\alpha_\mathrm{p}$    & 3.772  \\
        Petal taper rate parameter                            & $\beta_\mathrm{p}$     & 0.037         \\
        Petal total width                                     & $p_\mathrm{w}$         & 378 mm        \\
        Petal behind backshort (fixed)                        & $p_\mathrm{b}$         & 20 mm         \\
        Petal corner radius                                   & $r_\mathrm{p}$         & 27.5 mm      \\
        Balun microstrip taper length                         & $b_\mathrm{l}$         & 48.3 mm      \\
        Balun microstrip width (port side)                    & $m_{\mathrm{w},\mathrm{p}}$     & 0.6 mm       \\
        Balun microstrip width (short-to-petal side)          & $m_{\mathrm{w},\mathrm{s}}$     & 4.3 mm       \\
        Balun microstrip taper scale parameter (fixed)        & $\alpha_\mathrm{b}$    & $5.116\times 10^{-9}$ \\
        Balun microstrip taper rate parameter                 & $\beta_\mathrm{b}$     & 0.428         \\ \hline
        \end{tabular}\label{tab:dimensions}
        \end{threeparttable}
\end{table}

A microstrip balun is added over the untapered slot of each petal, as shown in shown in \autoref{subfig:baluns_on_feed}. The balun closest to the backshort is labeled ``polarization 0,'' and the one closest to the exponential taper is labeled ``polarization 1.'' Each balun is a double-sided PCB, consisting in a microstrip on one side, and a grounded shield on the other. They are raised 0.8\,mm above the petal, leaving an air gap---we call them \textit{floating} baluns---with the microstrip's side facing the petals. The floating baluns minimize losses in materials that would otherwise occur between the balun and the petal if that space was filled with a dielectric, as the electric field density is high in that volume as seen in \autoref{fig:currents}. The baluns shown in \autoref{fig:feed_alone} and \autoref{fig:baluns_fig} are terminated with a short---an off-the-shelf connection pin---to the petal at one end. At the other end, they can either be terminated with an SMA connector, or directly soldered to a custom amplifier to avoid insertion loss in the connector, and a castellated hole enables easy soldering. The value of 0.8\,mm for the air gap was chosen as it is the thickness of the custom CHORD amplifier (introduced in \cite{Mark_LNA}). We have explored thinner and thicker options but did not find significant improvements in either the loss or the impedance matching.

The balun's microstrip is also exponentially tapered, and its degrees of freedom are: its length, width at either end, and taper rate parameter $\beta_\mathrm{b}$---as was the case for the petal's taper, the balun taper scale parameter $\alpha_\mathrm{b}$ is not free, since it is determined by the taper width and length. The balun thickness is also fixed, at 0.8\,mm, because anything thinner would lack the structural integrity needed for the floating balun design, and anything thicker would necessitate the petal's untapered slot to be wider, pushing the feed's impedance away from the desired impedance. The choice of substrate for the balun affects both the impedance, and the losses in dielectrics. We found that it is hard to reach a 50\,$\Omega$ impedance across the band with teflon ($D_k\sim 2.1$, $\tan\delta\sim 0.0004$), while FR-4 ($D_k\sim 4.5$, $\tan\delta\sim 0.025$) is too lossy; it resulted in a noise temperature contribution of up to 14\,K at the higher end of the band (0.2\,dB loss). The baluns tested were thus printed on Rogers RO4003C, with $D_k=3.55$, $\tan\delta\sim 0.0025$, whose noise temperature contribution peaks at 1.4\,K (0.02\,dB loss), and allows for an optimal impedance match, both presented in \autoref{subsec:CHORD_receiver_performance}.

The values of all the parameters, optimized for the CHORD feed as presented in \autoref{sec:CHORD}, are shown in \autoref{tab:dimensions}.

\subsection{Optimization}
\label{subsec:optimization}

The success of the optimization process relies on the fact that array elements in ``large-$N$, small-$D$'' observatories are electrically small, such that with modern high performance computing facilities, simulations that include the dish converge relatively quickly. Without any hardware acceleration, a simulation takes less than 2 minutes to converge using a model of the feed without a dish. If we add a 6\,m dish, remove the feed baluns to take advantage of spatial symmetries, and enable far field monitors at a 0.1\,GHz frequency resolution, the simulation time only increases to approximately 8 minutes. This allows us to use algorithms that require a $\mathcal{O}(100)$ iterations, and exhaustively explore a vast parameter space, optimizing the dish beam properties directly without having to estimate them from simulations of the feed alone using physical optics approximations. Note that removing the baluns means that when simulating the feed in the dish, the impedance cannot be optimized for, hence, the impedance and the beam shape are optimized separately. A frequency-domain solver is used when simulating the feed alone, as it produced S-parameters that more closely matched measurements, and converged faster with an adaptive mesh algorithm. A time-domain solver is used when simulating the feed with a 6\,m dish to optimize the beam, because it was faster for that setup, and the beam pattern was the same with either solver.

Before the optimization begins, the feed dimensions are set to the following values: the total width and length of the petals at about a third of the longest wavelength, and just a bit less than half of that for the backshort diameter and taper length. Those dimensions were found through an exploration process that involved modeling many variations on Vivaldi feed models found in modern literature \citep{vivaldi_stubs_1,Vivaldi_slots1,Vivaldi_parasitic1,vivaldi_lumped_resistor_1}, with custom modifications and very vast parameter space. While aiming for the CHORD specifications over a 5:1 band---an aperture efficiency near or above 50\%, $T_\mathrm{sys}$ contributions below 30\,K, and a 50\,$\Omega$ impedance---trial and error led to the oversized backshort as the ideal option, a variation that we did not find in literature. Note that at this stage of the process, we did not precisely aim for the impedance of the custom CHORD LNA, as its design was not finalized. When we did obtain the LNA's $\Gamma_\mathrm{opt}$, which was close to 50\,$\Omega$, we did not change the basic feed design, and simply ran the optimization algorithm described below. We propose the same strategy for the use of this design on another observatory: starting from the approximate dimensions we found, and using the following algorithm to adapt the feed to specific impedance and beam shape goals.

The optimization process consists of three steps, which are performed twice. In the first step, the feed is optimized over all the petal and balun parameters to reach a desired impedance, using a covariance matrix adaptation evolution strategy (CMA-ES) algorithm, because CMA-ES performs well with a large number of varying parameters, far from the goal function minimum \citep{cma_es}.

In the second step, the full dish is added to the model, but the baluns are removed and symmetry planes parallel to the petals are added to limit the mesh resolution and simulation time. Another CMA-ES optimization run is performed, this time optimizing the beam shape. Depending on the specific science goals, this can mean a high forward gain, a low spillover, or another metric. In those simulations, the balun parameters are not considered, but one new parameter is added, that is, the feed position with respect to focus. Also, the parameters that were found in the first optimization run are kept within a narrower range.

In the third step, the dish is removed, the baluns are added back, and only the balun parameters are optimized over, to try and correct for the impedance changes caused by the previous step without affecting the beam shape. This time, however, the algorithm chosen is Nelder-Mead, which performs well over a small number of parameters \citep{nelder_mead}.

\begin{figure}[h!]
\centering
    \begin{tikzpicture}
        \node[terminator, fill={myblue1}] at (0,0) (start) {\textbf{Start}};
        \node[process, fill={myyellow1}] at (3,0) (A1) {\\\textbf{Optimize:}\\Impedance\\\textbf{Included:}\\Petals, baluns\\\textbf{Algorithm:}\\CMA-ES};
        \node[process, fill={myyellow1}] at (7.0,0) (B1) {\\\textbf{Optimize:}\\Beam shape\\\textbf{Included:}\\Petals, dish (fixed)\\\textbf{Algorithm:}\\CMA-ES};
        \node[process, fill={myyellow1}] at (11.5,0) (C1) {\\\textbf{Optimize:}\\Impedance\\\textbf{Included:}\\Petals (fixed), baluns\\\textbf{Algorithm:}\\Nelder-Mead};
        \node[process, fill={myyellow1}] at (3,-3.5) (A2) {\\\textbf{Optimize:}\\Impedance\\\textbf{Included:}\\Petals, baluns\\\textbf{Algorithm:}\\Trust-region};
        \node[process, fill={myyellow1}] at (7.0,-3.5) (B2) {\\\textbf{Optimize:}\\Beam shape\\\textbf{Included:}\\Petals, dish (fixed)\\\textbf{Algorithm:}\\Trust-region};
        \node[process, fill={myyellow1}] at (11.5,-3.5) (C2) {\\\textbf{Optimize:}\\Impedance\\\textbf{Included:}\\Petals (fixed), baluns\\\textbf{Algorithm:}\\N-M or T-r};
        \node[draw=none] at (15,0) (topright) {...};
        \node[draw=none] at (0,-3.5) (bottomleft) {...};
        \node[terminator, fill={myblue1}] at (15,-3.5) (end) {\textbf{End}};
        \path [connector] (start) -- (A1);
        \path [connector] (A1) -- (B1);
        \path [connector] (B1) -- (C1);
        \path [connector] (C1) -- (topright);
        \path [connector] (bottomleft) -- (A2);
        \path [connector] (A2) -- (B2);
        \path [connector] (B2) -- (C2);
        \path [connector] (C2) -- (end);
    \end{tikzpicture}
\caption{Overview of the optimization process. At each step, the range of the parameters are narrowed.} \label{fig:opt_flowchart}
\end{figure}
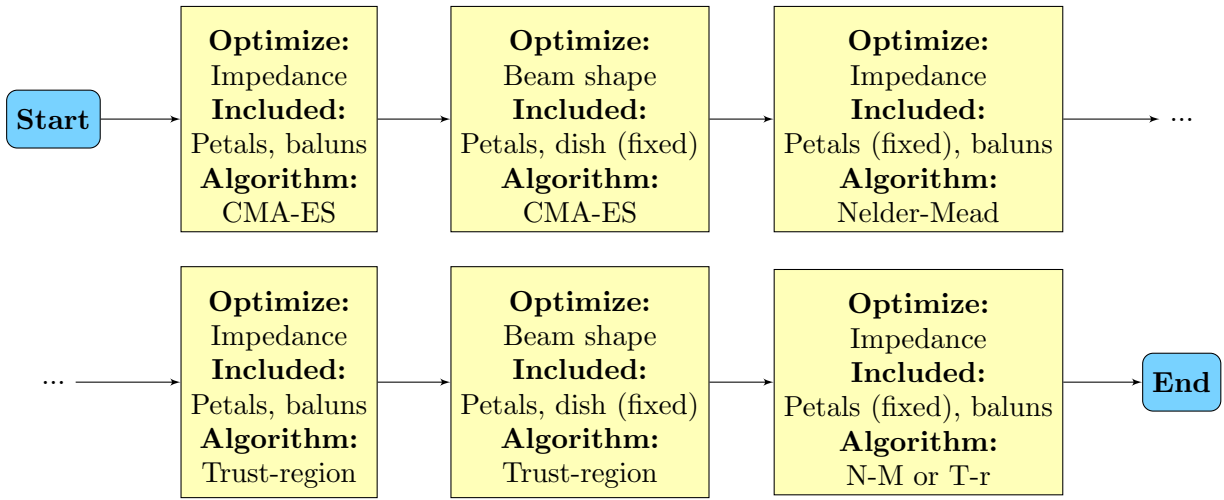

After those three steps are completed, the feed is considered to be close to the global optimum; the steps are then repeated with a slightly different strategy that is better adapted for a smaller parameter space. For the first two steps, the range of variation for the parameters is narrowed, and a trust-region framework algorithm is used for optimization, which performs well when near a goal function minimum \citep{trust_region}. The third step is also repeated, and can be performed with either the Nelder-Mead (because there are only a few parameters involved) or trust-region framework (due to proximity to goal function minimum) algorithm---both taking similar numbers of iterations to converge, at that point in the process. At the end, a full simulation that includes both the dish and the baluns should be performed, to check that the performance is as desired. This process is summarized in \autoref{fig:opt_flowchart}.

\subsection{Manufacture}
Once the feed model meets the performance goals in simulations, some details are added for practical reasons. For instances, holes are added on the petals and in the backplane, to make room for screws, connectors, cables, and first-stage amplification. A piece of polycarbonate is also inserted in the exponential taper, for stabilization purposes. The piece is 3\,mm in thickness, contributing to less than 1\,K to the system temperature (0.014 dB loss) over the band. Simulations indicate that the presence of this dielectric in the taper does not affect the beam shape or the impedance. Like the aluminum, it is laser cut, adding no complexity to the manufacture. Once the parts are laser cut, and the baluns are printed, everything is assembled with rivets or screws. The resulting assembled feed is shown in \autoref{fig:feed_alone}.

\section{Performance}
\label{sec:performance}

The feed that was built and tested was optimized for the CHORD observatory, as described in \autoref{sec:CHORD}. It is made to match the custom CHORD first-stage LNA's optimum impedance for minimum noise, and to have a beam that meets the constraints set by CHORD's science goals. However, since we argue that this feed design could be used for other ``large-$N$, small-$D$'' arrays, we present more generic benchmarks---the reflection coefficient referenced to 50\,$\Omega$, beam shape, aperture efficiency, contribution to $T_\mathrm{sys}$ from spillover at various focal ratios, and cross-polarization. They already meet standard benchmarks, and can be improved once optimized for a given observatory.

\begin{figure}[h!]
    \centering
    \includegraphics[width=\textwidth]{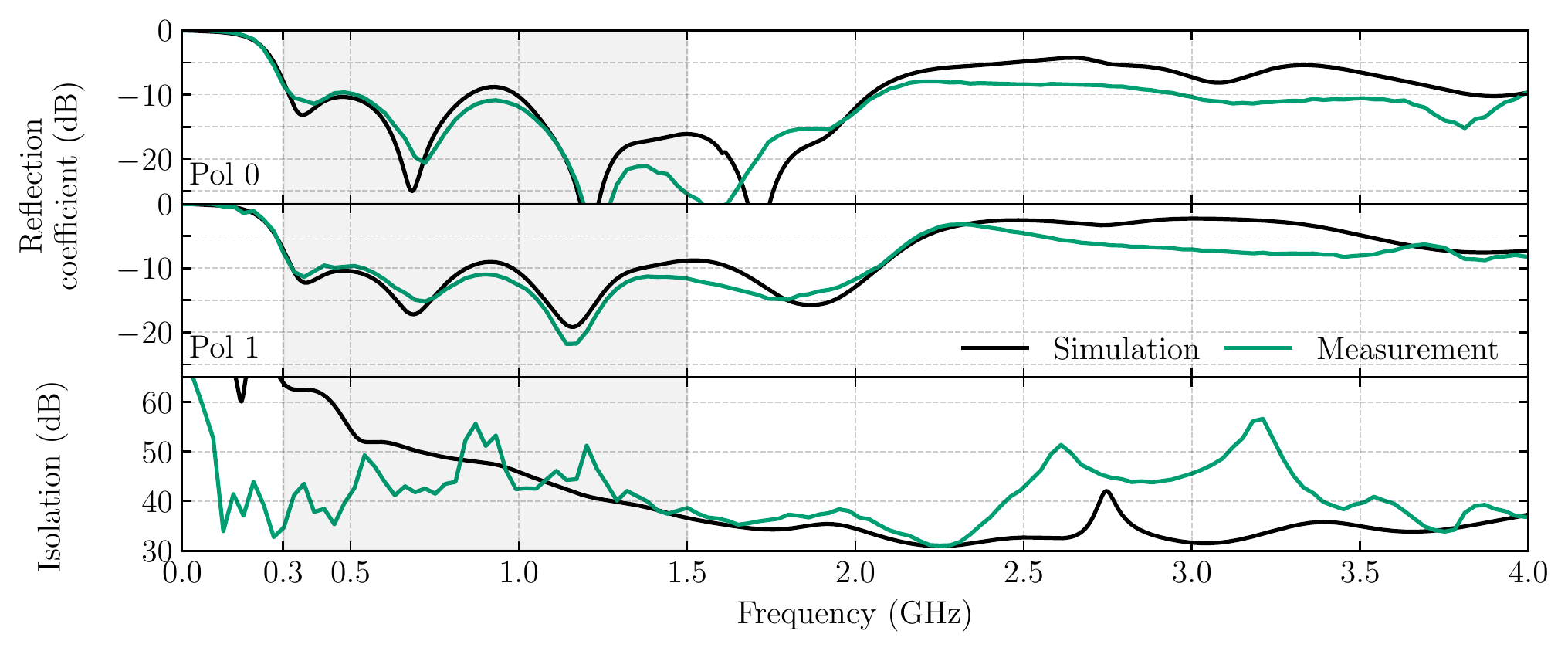}
    \caption{Reflection coefficient from DC to 4\,GHz (CHORD band shaded), for both polarizations, in simulation, and measured with a vector network analyzer connected to the baluns through SMA connectors.}
    \label{fig:reflection}
\end{figure}

The reflection coefficient of the feed is measured at the SMA-end of the SMA-terminated balun (as shown in \autoref{fig:baluns_fig}) with a vector network analyzer that was calibrated to the reference plane of the SMA. The simulated and measured reflection coefficients for both polarizations, referenced to 50\,$\Omega$, are presented in \autoref{fig:reflection}. Measurements match simulation, and both are below $-10$\,dB over the CHORD band (0.3--1.5\,GHz) averaging around $-16$\,dB for polarization 0, and $-13$\,dB for polarization 1. The reflection coefficient almost even clears the $-10$\,dB threshold over a 7:1 band, but beam properties were not optimized above 1.5\,GHz. The isolation between both polarizations is above 30\,dB over the whole measured and simulated range.

\begin{figure}[h!]
     \centering
     \includegraphics[width=\textwidth]{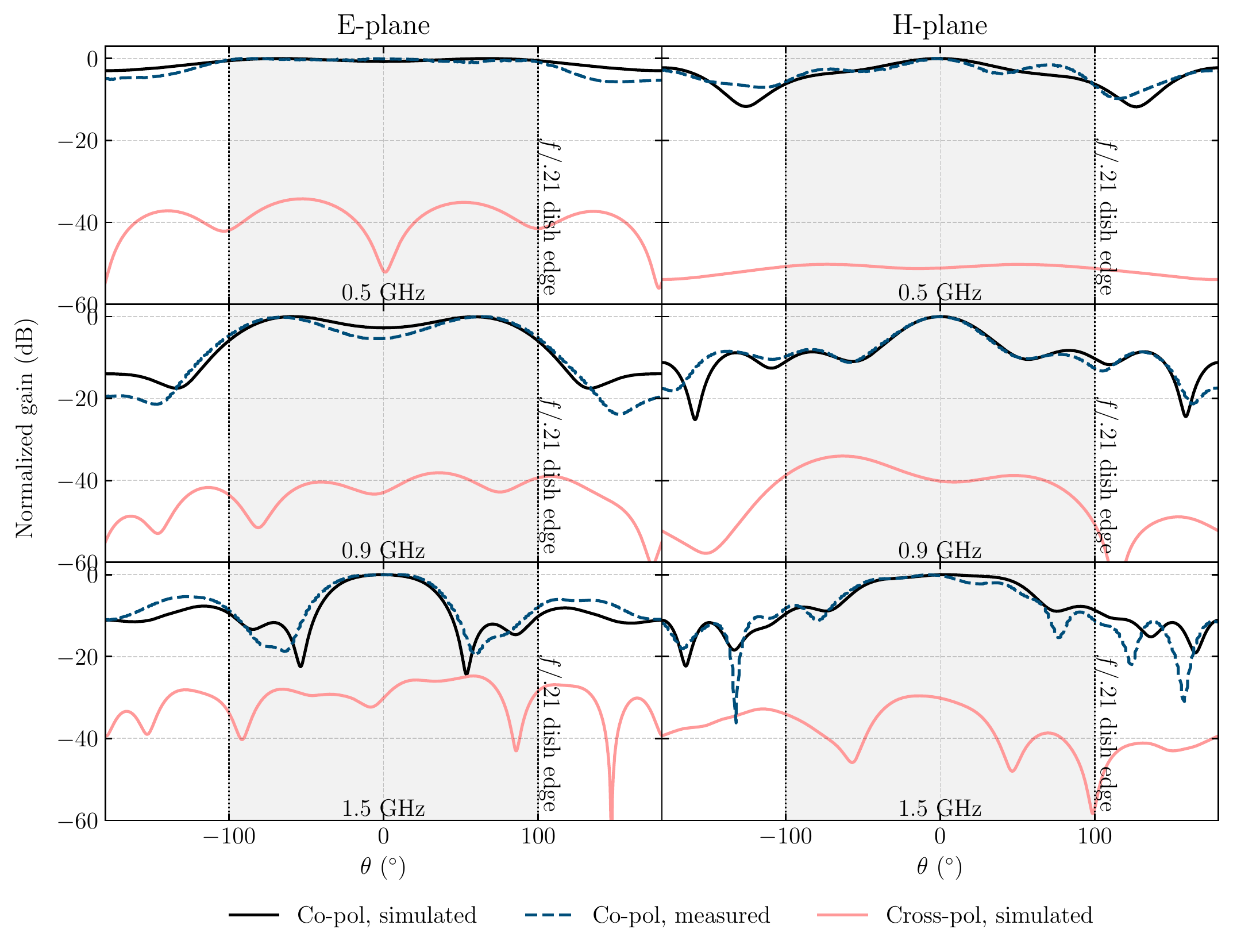}
     \caption{E-field beams of the feed alone at 0.5, 0.9, and 1.5\,GHz, from simulations and anechoic chamber measurements. The chamber was not rated for measurements below 0.5\,GHz, or for reliable cross-polarization measurements.}
    \label{fig:feed_beams}
\end{figure}

The E-field beams of the feed alone---in CST simulations and anechoic chamber measurements---are presented in \autoref{fig:feed_beams} at 0.5, 0.9, and 1.5\,GHz, along the E-plane and H-plane. Frequencies below 0.5\,GHz are not presented as the anechoic was not rated for such long wavelengths. In the E-plane cut, the low frequency beam is nearly isotropic, peaking near $\theta=60^\circ$ in simulations. Those very wide beams are well suited for deep dishes: most of the power near $\theta\sim90^\circ$ is captured by the dish when $f/D\leq 0.25$, and the portion that spills over is radiated towards the sky rather than the ground, when pointed near zenith---in that setup, the only radiation that reaches the ground is that diffracted off the edge of the dishes.

\begin{figure}[h!]
    \centering
    \includegraphics[width=\textwidth]{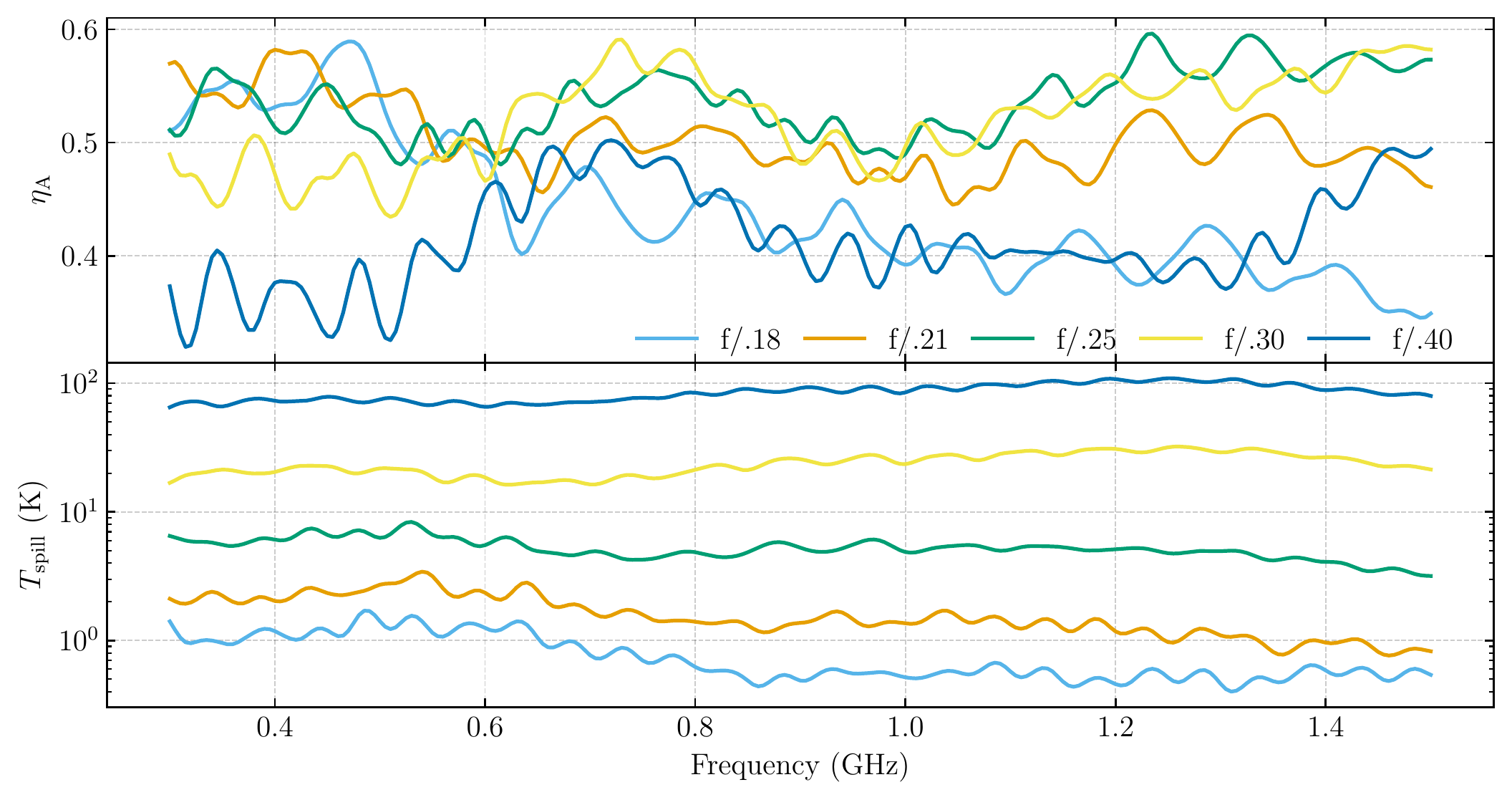}
    \caption{(Top) Simulated aperture efficiency of the feed when placed in dishes with various focal ratios. (Bottom) Simulated contributions to $T_\mathrm{sys}$ from spillover when the feed is placed in those same dishes of varying depths, when pointed at zenith. Contributions at higher zenith angles are presented in \autoref{fig:Tspill_Tsys}. The frequency resolution is 0.005\,GHz, and the simulations include the dish and the feeds, but excludes any support structure.}
    \label{fig:eta_A_T_sys}
\end{figure}

\autoref{fig:eta_A_T_sys} shows the trade-off between illumination and spillover efficiency at different focal ratios, from simulations that included the feed and the dish only (no support structure was modeled). The $\sim 0.12$\,GHz ripple is due to reflections between the feed and the dish vertex---mitigation efforts are underway, not covered in this paper. Both the illumination and the spillover start decreasing at higher focal ratios such that the feed's performance is poor when used on shallow dishes ($f/D\geq 0.3$). The focal ratio where the average aperture efficiency over the band is maximized, at $\bar{\eta}_\mathrm{A}=0.54$, is $f/D=0.25$. We note that since the beam is so wide at low frequencies, the dish remains well illuminated at that end of the band, even down to $f/D = 0.18$. This results in a downward slope in $\eta_\mathrm{A}$ over frequency, which could be used to maintain a more constant beam width over the band. That may be desirable, for instance, to reduce foreground leakage in 21\,cm experiments \cite{beamwidthforegrounds_alonso}. The aperture efficiency is not linear with frequency or focal ratio since the feed beam's power taper is not smooth along either of those axes, as can be seen in \autoref{fig:feed_beams}. The optimal dish depth ultimately depends on the specific science goals and constraints of a given observatory and, as explained in \autoref{sec:CHORD}, the CHORD dishes were chosen to be have a focal ratio of $f/D=0.21$. The aperture efficiency is nearly as high as $f/D = 0.25$, but the ground illumination contributes $5\sim7$\,K less to $T_\mathrm{sys}$. Note that spillover was not the only factor in favor of a lower focal ratio: mitigation of crosstalk, as mentioned in \autoref{subsec:CHORD_crosstalk}, was also considered. In simulations with multiple dishes, we found that going from $f/D=0.25$ to $f/D=0.21$ significantly reduced adverse coupling effects as the feeds are not in line-of-sight of each other anymore. The details of those simulations are beyond the scope of this paper, and are yet only published in an internal CHORD memorandum---the HIRAX collaboration made similar findings, presented in \cite{HIRAX_overview}.

\begin{figure}[h!]
    \centering
    \includegraphics[width=\textwidth]{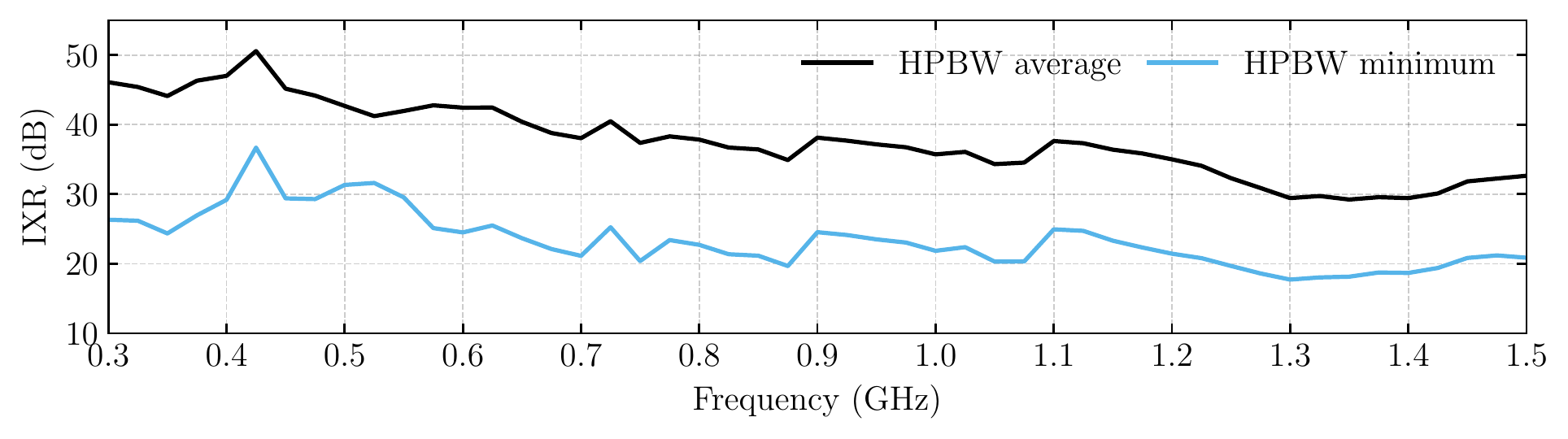}
    \caption{Average and minimum intrinsic cross-polarization ratio (IXR) within the dish beam's HPBW, simulated with a 6\,m reflector with $f/D=\,0.21$ fed with the feed design.}
    \label{fig:crosspol_IXR}
\end{figure}

\autoref{fig:crosspol_IXR} shows the minimum and average intrinsic cross-polarization ratio (IXR) within the half-power beam width (HPBW), simulated with a 6\,m reflector with $f/D=\,0.21$ fed with the feed design. The IXR is a figure of merit for cross-polarization introduced in \citealt{IXR} specifically for radio astronomy purposes, that is independent of the coordinate system, contrary to the cross-polarization isolation (XPI) or discrimination (XPD). It is equal to
\begin{equation}\label{eq:IXR}
    \mathrm{IXR} = \left|\frac{\sigma_\mathrm{max} + \sigma_\mathrm{min}}{\sigma_\mathrm{max} - \sigma_\mathrm{min}}\right|^2,
\end{equation}

where $\sigma_\mathrm{max}$ and $\sigma_\mathrm{min}$ are the maximum and minimum singular values of the dish beam's Jones matrix. We are not aware of an standard benchmark for that metric, but large-$N$ astrophysical observatories such as the SKA have quoted $\mathrm{IXR}\geq 15$\,dB everywhere within the HPBW as a specification, which this feed design clears \cite{SKA_IXR}.

\begin{figure}[h!]
    \centering
    \includegraphics[width=\textwidth]{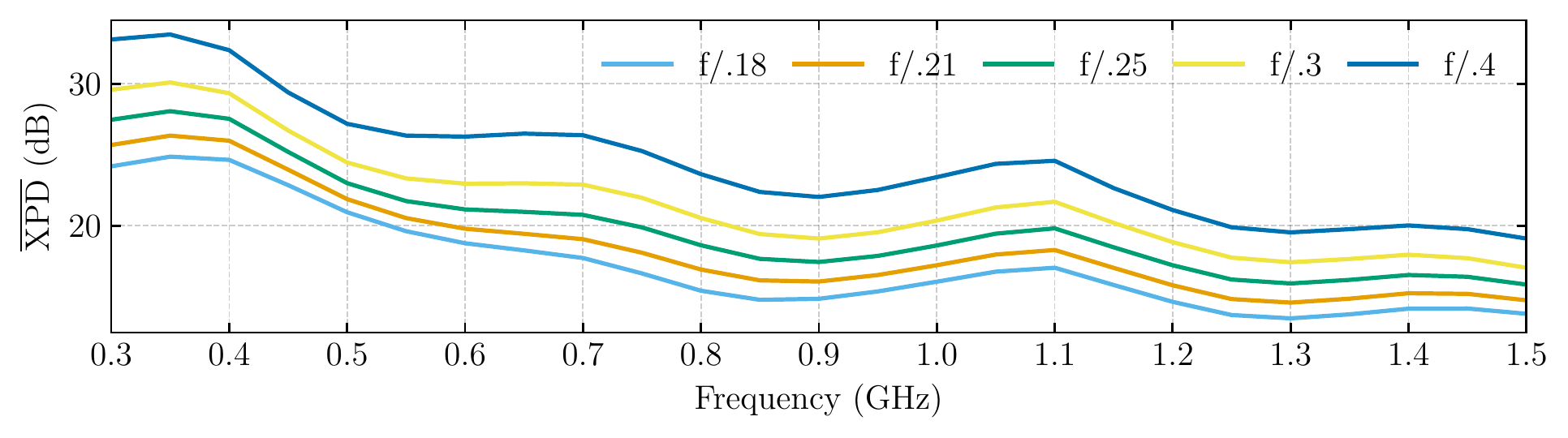}
    \caption{Average cross-polarization discrimination (XPD) of the feed beam within the angle subtended by a dish of focal ratios 0.18, 0.21, 0.25, 0.30, and 0.40, in simulations.}
    \label{fig:XPD}
\end{figure}

In \autoref{fig:XPD}, the average cross-polarization discrimination (XPD) within the angle subtended by a dish of focal ratios 0.18, 0.21, 0.25, 0.30, and 0.40 is presented. As the dish gets deeper, the average XPD gets lower, suggesting that the radiation from the feed has higher cross-polarization levels towards the edges than close to boresight. CHORD does not have an XPD specification, as its main science goals do not primarily depend on a high XPD.

\section{Implementation on CHORD}\label{sec:CHORD}

\subsection{CHORD overview}
\label{subsec:CHORD_overview}
CHORD's main core will be localized at the Dominion Radio Astrophysical Observatory (DRAO), near Penticton, in British Columbia, Canada. It will be comprised of 512 closely packed 6\,m prime-focus dishes, distributed in a regular rectilinear array, with East-West and North-South shortest baselines of 6.3\,m and 8.5\,m, respectively. The dishes will have an inclination range of 30$^\circ$ from zenith, in the North-South direction only. The bandwidth of its receiver will be 0.3 to 1.5\,GHz \cite{CHORD_overview}.

The science and engineering considerations that lead to those design decisions are not the subject of this paper, but a quick summary of those relevant to the optimization of the feed design are presented.

\subsection{Crosstalk}
\label{subsec:CHORD_crosstalk}
The very short 6.3\,m baselines---corresponding to a $\lambda_0/3$ gap from the edge of one dish to the next---are determined by the cosmology goals, as large scales of the matter power spectrum are only accessible with short baselines in cross-correlation. This proximity increases the mutual coupling (crosstalk) between array elements, which has the undesired effect of correlating noise radiated out of the receivers, or picked up from the ground. Moreover, crosstalk causes the sky signal to be reflected from one array element and detected by another, adding chromaticity to the measurement. This chromaticity complicates the process of subtracting galactic foregrounds, which makes use of their otherwise smooth spectrum. Calibrating the instrument to take crosstalk into account in analysis is an active area of research (see \citealt{crosstalk_kern_1}, \citealt{crosstalk_kern_2}, and \citealt{crosstalk_josaitis}) but we can mitigate that issue by minimizing the coupling power in the design phase. Additionally, due to the faintness of the low redshift 21\,cm signal---$\sim 0.1$\,mK brightness temperature---the constraints on noise are very strict, at $T_\mathrm{sys}<30\,K$ across the band, with ground illumination being a large contributor. One way to reduce both crosstalk and spillover in other observatories has been by engineering the optics of the reflectors, for instance, by using carefully shaped sub-reflectors and shields \citep{optics_allen,optics_meerkat,optics_ska,optics_ngVLA}. However, due to the cost and complexity constraints, those approaches are not possible with CHORD, which will use prime-focus dishes. Instead, the CHORD dishes minimize crosstalk and spillover by being very deep, with $f/D=0.21$, the edges of the dish thus effectively acting as a shield.

\subsection{Optimization for CHORD}
\label{subsec:CHORD_optimization}
To measure the 21\,cm signal, minimizing $T_\mathrm{sys}$ takes precedence over the conventional $T_\mathrm{sys}/\eta_\mathrm{A}$. The reasoning is that the benefits of a higher $\eta_\mathrm{A}$ are mitigated by the reduced field of view when measuring an extended source whose features are large in scale, while $T_\mathrm{sys}$ directly increases the SNR without any drawback \citep{Morales2005Feb}. As such, minimizing $T_\mathrm{sys}/\eta_\mathrm{A}$ could marginally benefit other science goals that require high point-source sensitivity, while negatively impacting 21\,cm science. Moreover, the feed's $T_\mathrm{sys}$ is dominated by the LNA and balun, which are independent of $\eta_\mathrm{A}$. The optimization process thus minimizes $T_\mathrm{sys}$ by simulating only the feed with baluns, and then maximizes $\eta_\mathrm{A}$ near the resulting solution, with balun parameters fixed. Lastly, in addition to maximizing $\eta_\mathrm{A}$, backward hemisphere radiation is minimized to decrease spillover and systematics such as crosstalk. The corresponding goal functions are summarized below.

The impedance goal function is simply the noise figure $F$, given by:
\begin{equation}\label{eq:noise}
    F = F_\mathrm{min} + \frac{4 R_\mathrm{n}}{Z_0}\frac{\left|\Gamma_\mathrm{S}-\Gamma_\mathrm{opt}\right|}{\left(1-\left|\Gamma_\mathrm{S}\right|^2\right)\left|1+\Gamma_\mathrm{opt}\right|^2},
\end{equation}
where $F_\mathrm{min}$ is the LNA's minimum noise figure, $R_\mathrm{n}$ is its equivalent noise resistance, and $Z_0$ is the reference impedance. The target is $F = 0$. \autoref{eq:noise} is taken from \citealt{ludwig_rf_2000}. Once a minimum is found, the optimizer looks for an optimal beam in the vicinity of this solution. The baluns are removed and the dish is 
added. The aperture efficiency $\eta_A$ is given by:
\begin{equation}\label{eq:eta_A}
    \eta_\mathrm{A} = \frac{A_\mathrm{eff}}{A_\mathrm{phys}},\quad\text{where}\quad A_\mathrm{eff} = \frac{\lambda^2 G}{4\pi},
\end{equation}
with $\lambda$, the wavelength, $G$, the gain at boresight, and $A_\mathrm{phys}$, the size of the physical aperture $\pi R^2$, $R=3$\,m. Its target is $\eta_A=1$. The ground illumination goal functions represent the integrated power pattern beyond some critical angle $\theta_\mathrm{c}$:
\begin{equation}
    P(\theta_\mathrm{c}) = \int_{0}^{360^\circ}\int_{\theta_\mathrm{c}}^{180^\circ} U(\theta,\phi)\sin(\theta)\mathrm{d}\theta\mathrm{d}\phi,
\end{equation}
where $U(\theta,\phi)$ is the normalized power pattern of the dish. The targets are $P(90^\circ)=0$ and $P(60^\circ)=0$.

During the noise matching optimization, the distance from the goal function to its target is minimized by the algorithms mentioned in \autoref{subsec:optimization}. During the beam optimization, its the weighted sum of the distances between each goal function and their respective target that is minimized with the same algorithms. The weights are: 1 for $\eta$ is unity, 0.5 for $P(90^\circ)$, and 0.2 for $P(60^\circ)$. Those values are chosen according to the following heuristic reasoning: first, $P(90^\circ)$ and $P(60^\circ)$ overlap, such that any fraction of the power beam beyond $60^\circ$ is by definition also beyond $90^\circ$. Then, $P(60^\circ)$ is downweighted even more as the dish only picks up ground radiation near $\theta=60^\circ$ when it is tilted at a $30^\circ$ zenith angle, which will be the case for a smaller fraction of CHORD's operations. The dimensions that the algorithm converged to are presented in \autoref{tab:dimensions}.

\subsection{CHORD receiver performance}
\label{subsec:CHORD_receiver_performance}
\begin{figure}[h!]
     \centering
     \includegraphics[width=0.5\textwidth]{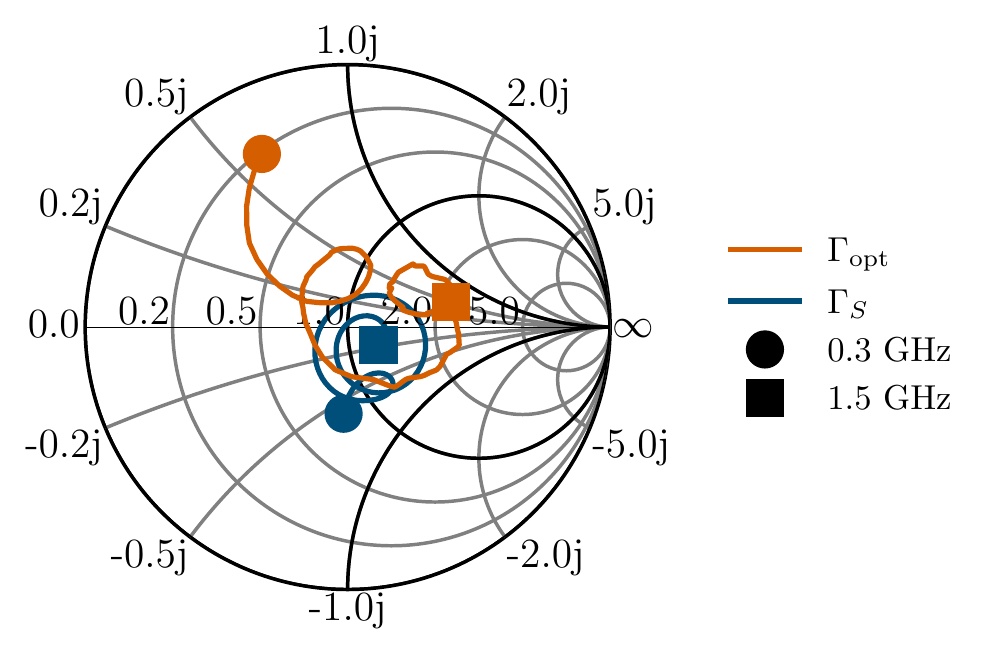}
     \caption{$\Gamma_\mathrm{opt}$ and $\Gamma_\mathrm{S}$, on a Smith chart referenced to 50\,$\Omega$. Both remain close to 50\,$\Omega$ over most of the band, except for $\Gamma_\mathrm{opt}$ near 0.3\,GHz. This partly explains the sharp increase in the LNA noise match contribution to $T_\mathrm{sys}$ at the very low end of the band, in \autoref{fig:Tspill_Tsys}.}
    \label{fig:Gammas}
\end{figure}

The feed was optimized for the CHORD constraints and hardware. The 5:1 CHORD bandwidth makes the LNA noise matching challenging, as the LNA's $\Gamma_\mathrm{opt}$ exhibits a negative-capacitance response, which is not possible to realize with Foster matching networks over the full bandwidths \citep{Belostotski_2012}. The approach taken for the CHORD LNA was to bring $\Gamma_\mathrm{opt}$ close to $50\thinspace\Omega$ by employing a combination of wideband matching at the LNA input and by using the intrinsic feedback through the parasitic gate-drain capacitance of the input stage transistor \citep{Zailer_2020,Beaulieu_2016,Belostotski_2007}.
The noise matching of the LNA can be understood from \autoref{fig:Gammas}, where a Smith chart, referenced to 50\,$\Omega$, plots both $\Gamma_\mathrm{opt}$ and $\Gamma_\mathrm{S}$. Both the LNA and the feed maintain their $\Gamma_\mathrm{opt}$ and $\Gamma_\mathrm{S}$, respectively, near 0 (i.e. 50\,$\Omega$) over the band, except at the very low end for the LNA \citep{Mark_LNA}.

\begin{figure}[h!]
    \centering
    \includegraphics[width=\textwidth]{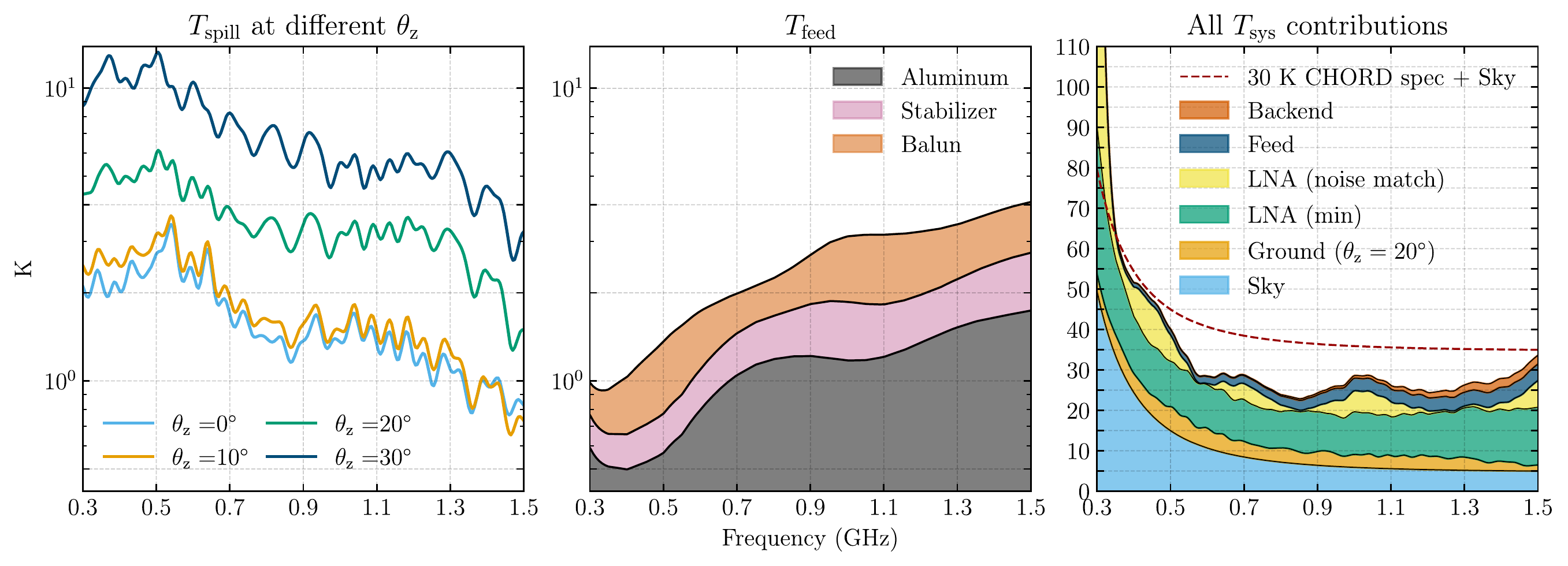}
    \caption{(Left) Contributions to $T_\mathrm{sys}$ from spillover at increasing zenith angles $\theta_\mathrm{z}$. (Center) Contributions to $T_\mathrm{sys}$ from the feed itself, from simulations. (Right) All contributions to $T_\mathrm{sys}$, assuming $\theta_\mathrm{z}=20^\circ$.}
        \label{fig:Tspill_Tsys}
\end{figure}

The leftmost plot of \autoref{fig:Tspill_Tsys} shows the contribution to $T_\mathrm{sys}$ from the ground spillover ($T_\mathrm{spill}$) at increasing zenith angles $\theta_\mathrm{z}$, in CHORD's range of $0^\circ$\,$<$\,$\theta_\mathrm{z}$\,$<$\,$30^\circ$. It is calculated by multiplying the ambient temperature by the fraction of the beam power reaching below horizon, $T_\mathrm{ground}=300\cdot P_\mathrm{ground}$. The spillover temperature is in the 2--3\,K range when the dishes are pointed within 10$^\circ$ of zenith, but goes up to 10--14\,K at the highest tilt ($\theta_\mathrm{z}=30^\circ$). The middle plot shows the contributions due to losses in the feed materials: the aluminum (0.5--1.7\,K), the polycarbonate stabilizer piece (0.1--1\,K), and the balun substrate (0.3--1.3\,K). The total on-sky system temperature $T_\mathrm{sys}$ with its individual contributions are shown on the right plot. The sky temperature is simplified as isotropic, and is modeled as 
\begin{equation}\label{eq:Tsky}
T_\mathrm{sky} = P_\mathrm{sky}\left(T_\mathrm{CMB}+T_{408}\left(\nu/\nu_{408}\right)^\beta + T_\mathrm{atm}\tau_\mathrm{z}\sec(\theta_\mathrm{z})\right),
\end{equation}
where $P_\mathrm{sky}$ is the fraction of the beam power reaching above horizon; the temperature of the cosmic microwave background $T_\mathrm{CMB}=2.73$\,K is taken from \cite{T_cmb}; the 408\,MHz galactic continuum temperature $T_{408}=20$\,K and spectral index $\beta=2.7$ are chosen as typical values away from the galactic plane from surveys \citep{Tsky_Haslam,Tsky_spectral}; the atmospheric temperature is set to the ambient temperature $T_\mathrm{atm}=300$\,K; the zenith opacity $\tau_\mathrm{z}$ is derived from models of the absorption features of oxygen \citep{Rosenkranz_1975} and water \citep{ulaby_microwave_2014} at sea level; $\theta_\mathrm{z}$ is the zenith angle and $\sec(\theta_\mathrm{z})$ is averaged to 1.1 since CHORD has $0^\circ$\,$<$\,$\theta_\mathrm{z}$\,$<$\,$30^\circ$. The sky temperature is of the order of 1\,K at the high end of the band, but reaches nearly 50\,K at low frequencies. The LNA noise temperature is computed with \autoref{eq:noise}, using the feed impedance and the LNA's optimal noise-matching impedance, as presented in \autoref{fig:Gammas}. The minimum noise temperature of the LNA is in the 10--15\,K range over the band, while the impedance mismatch adds 5--7\,K at specific frequencies; both increase dramatically to near 250\,K in the lower 20\,MHz of the band. The entire receiving chain operates at ambient temperature.

The system temperature meets the CHORD specification of $T_\mathrm{sys}<30$\,K over nearly all the band, apart from the lower end where the sky gets too bright and the LNA's noise temperature increases appreciably. One significant contributor to the noise temperature is the spillover, which competes directly with aperture efficiency $\eta_\mathrm{A}$: to reduce ground illumination, the feed's beam needs to be more directed, which under-illuminates the dish and reduces forward gain. Varying the feed dimensions allows for a fine-tuning of that trade-off, and the final design was chosen to maintain $T_\mathrm{sys}$ at specification values, while keeping $\eta_\mathrm{A}$ near 50\%. The simulated aperture efficiency on the CHORD dishes is presented in \autoref{fig:eta_A_T_sys}, as the $f/D = 0.21$ line. No contribution from the SMA insertion loss is included in \autoref{fig:Tspill_Tsys}. Indeed, when this feed is used on CHORD to meet the very strict noise temperature requirement, custom LNA is soldered directly to the balun, as shown in \autoref{subfig:baluns_on_feed}, thus avoiding SMA insertion loss before amplification

\begin{figure}[h!]
     \centering
     \includegraphics[width=1.0\textwidth]{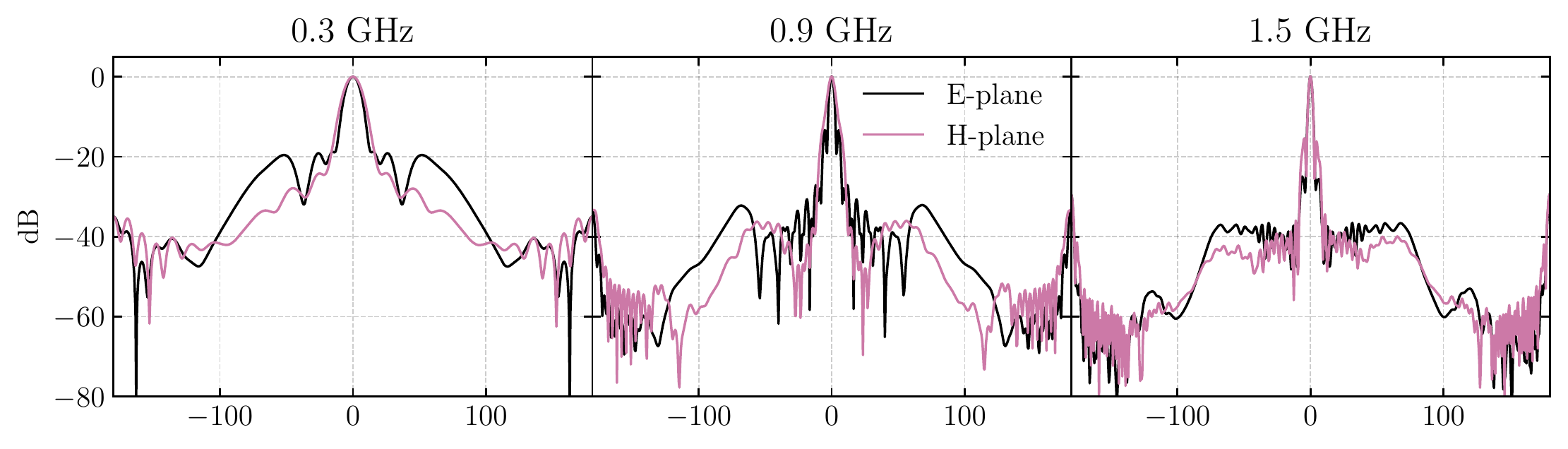}
     \caption{Simulated CHORD dish (6\,m, $f/D = 0.21$) radiation patterns when fed with the proposed feed, at 0.3, 0.9, and 1.5\,GHz, along three azimuthal cuts. The wide beams at low frequencies shown in \autoref{fig:feed_beams} are mostly captured by the very deep dish, leading to narrow main beam, high forward gain, and relatively low sidelobe levels ($<-$20\,dB).}
        \label{fig:dish_beams}
\end{figure}

The simulated CHORD dish beams are presented in \autoref{fig:dish_beams}. As explained in \autoref{sec:performance}, despite the very wide feed beams, the dish beams have narrow main lobes since the feed beams are captured by the deep $f/D=0.21$ dishes. The side lobe levels remains below $-20$\,dB across the band.

The feed blockage was estimated by running physical optics (PO) simulations with the far field beam of the feed inside of an $f/D=0.21$ dish, and comparing the resulting aperture efficiency with that of the full system simulation. The result is 0 to 5\% blockage across the band, with an average of 2\%.

\section{Conclusion}

We have developed a feed design that is well-suited for ``large-$N$, small-$D$'' observatories, especially when the dishes are very deep ($f/D\lesssim 0.25$). The feed assembly is fast and simple ($\sim$~20 minutes) and inexpensive to manufacture ($\lesssim$~75\,USD). It is made of laser cut aluminum, with compact PCB baluns, such that it exhibits low losses in materials (0.014--0.058\,dB). It achieves a 5:1 frequency ratio while remaining small ($\lesssim 0.4\times0.3\lambda_0$), thanks to an innovation: increasing the size of the backshort. It meets standard benchmarks in impedance and aperture efficiency when fed into very deep dishes ($f/D \leq 0.25$), and operates at ambient temperature. We present a strategy to efficiently optimize both impedance and beam properties using different algorithms depending on the number of free parameters and on the distance to the goal function minimum.

A realization of this feed design was built after being optimized for the CHORD observatory, a next-generation observatory set to be built in British Columbia, Canada, composed of 512\,$\times$\,6\,m dishes, arranged in a regular rectilinear array. The feed meets the CHORD's specifications, with a low system temperature ($T_\mathrm{sys}<30$\,K) for 21\,cm power spectrum measurements, and a high aperture efficiency for faint transient search.

\section{Acknowledgements}

CHORD is funded by the Canada Foundation for Innovation (CFI) 2020 Innovation Fund, and by contributions from the provinces of Alberta, Ontario, and Québec. We acknowledge the support of the Natural Sciences and Engineering Research Council of Canada (NSERC) grants: NSERC Discovery, and NSERC Discovery Accelerator. This research was also undertaken, in part, thanks to funding from the Canada Research Chairs Program. We would like to acknowledge CMC Microsystems and Canada’s National Design Network (CNDN) for the provision of Keysight ADS and Altium Designer that facilitated this research.

\clearpage
\newpage
\bibliographystyle{ws-jai}
\bibliography{bibliography}
\end{document}